\documentclass{article}
\usepackage{amssymb,amsmath,amsthm,amscd,graphicx,epsfig,latexsym}

\def\Ll{{\Lambda}  }
\def\G{{\Gamma}  }

\def\R{{\mathbb{R}}}
\def\C{{\mathbb{C}}}
\def\Z{{\mathbb{Z}}}
\def\Z2{\mathbb{Z}_2}

\def\T{\mathbb{T}  }
\def\V{\mathcal{V} }
\def\D{\mathbb{D} }

\def\LlF{(\D,\V,\T)  }

\def\EA{\Sigma }      

\def\EAL{{\EA}_{\Ll}  }
\def\EALa{{\EA}_{\Ll_1}  }
\def\EALaa{{\EA}_{\Ll_2}  }
\def\EALi{{\EA}_{\Ll_i}  }
\def\EALj{{\EA}_{\Ll_j}  }
\def\EAP{{\EA}_{\Pi}  }

\def\muL{\mu_{\Ll}   }
\def\P{\mathbb{P}  }
\def\f{\phi}            

\def\fla{A^*_{\Ll}  }
\def\fpa{A^*_{\Pi}  }
\def\MCE{{\EA^{*}}  }        
\def\S{\Omega }     
\def\H{(\S,\EA,\mu) }     
\def\HL{(\Ll,\EAL,\muL) }
\def\HP{(\S,\EA,\P) }

\def\B{\mathcal{B}  }

\def\vV{\triangle V  }

\def\sO{\mathcal{O}  }
\def\sV{{R_D }  }

\def\Cl{{\cal{C}}(\Omega,\EA,\mu)} 
\def\ClL{{\cal{C}}(\Ll,\EAL,\mu |_{\Ll})} 
\def\M{{\cal{M}}(\Omega,\EA,\mu)} 

\newtheorem{theorem}{Theorem}

\newtheorem{lemma}{Lemma}
\newtheorem{corollary}{Corollary}

\newtheorem{definition}{Definition}

\theoremstyle{remark}
\newtheorem{example}{Example}[section]

\newcommand\beq{\begin{equation}}
\newcommand\eeq{\end{equation}}
\newcommand\bea{\begin{eqnarray}}
\newcommand\eea {\end{eqnarray}}
\newcommand\ba{\begin{array}}
\newcommand\ea {\end{array}}

\newcommand\bd{\begin{description}}
\newcommand\ed {\end{description}}
\newcommand\ben{\begin{enumerate}}
\newcommand\een{\end{enumerate}}

\newcommand\bD{\begin{definition} }
\newcommand\eD{\end{definition} }
\newcommand\bE{\begin{example} }
\newcommand\eE{\end{example} }
\newcommand\bL{\begin{lemma} }
\newcommand\eL{\end{lemma} }
\newcommand\bT{\begin{theorem} }
\newcommand\eT{\end{theorem} }

\newcommand\bC{\begin{corollary} }
\newcommand\eC{\end{corollary} }

\def\nn{{\nonumber} }

\begin{document}

\vfill


\vfill
\vfill

\begin{center}
   \baselineskip=16pt
   \begin{LARGE}
      \textsl{Consistent Histories as Valuations}
   \end{LARGE}
   \vskip 2cm
      Yousef Ghazi-Tabatabai
   \vskip .6cm
   \begin{small}
      \textit{yousef.ghazi05@imperial.ac.uk}
        \end{small}
\end{center}

\vskip 1cm

\begin{small}
\begin{center}
   \textbf{Abstract}
\end{center}
Sorkin's \emph{coevent interpretation} \cite{Sorkin:coeventsI,Sorkin:coeventsII} shifts the focus of quantum logic from the structure of a propositional lattice to the nature of truth valuations thereon. We apply this shift in emphasis to a simple formulation of the consistent histories approach, expressing it in terms of truth valuations which are brought together in a \emph{logical framework} \cite{Yousef:CoeventsAsBeables}. We see that these consistent histories valuations are related to Sorkin's multiplicative coevents, and that they can be naturally described by an adaptation of Isham's early topos-theoretic approach \cite{Isham:1997}.
\end{small}
\vskip 0.5cm

\section{Introduction}\label{sec:introduction}

\subsection{Opening Comments}

The histories approach to quantum mechanics began with the spacetime paths of Dirac \cite{Dirac:1933} and later Feynman \cite{Feynman:1948,Feynman:1965}, and was subsequently generalised by Omnes \cite{Omnes:1988ek}, Hartle \cite{Hartle:1992as} and Gell-Mann \cite{Gell:1990}. More recently, it has been rephrased as a generalisation of probability theory in Sorkin's quantum measure theory \cite{Sorkin:1994dt,Sorkin:1995nj}.

However the interpretation has proved more elusive than the dynamics, leading to a variety of lines of inquiry. Perhaps the most prominent of these is the \emph{consistent histories interpretation} \cite{Griffiths:1984rx,Griffiths:1993,Griffiths:1996,Griffiths:1998}, the topos based analysis of which \cite{Isham:1997} has led to a topos-theoretic approach to quantum mechanics \cite{ Isham:1997,Isham:toposI,Isham:toposII,Isham:toposIII,Isham:toposIV,Butterfield:2003,Flori:2008fb,Isham:1999kb}. Alternatively, Sorkin's \emph{coevent interpretation} \cite{Sorkin:coeventsI,Sorkin:coeventsII,Sorkin:2010,Yousef:CoeventDynamics,Yousef:KSP,Yousef:thesis,Gudder:2009a,Gudder:2009b,Gudder:2010,Yousef:CoeventsAsBeables} has focused on truth valuation maps rather than the structure of the propositional lattice.

\subsection{Goals and Outline of this Paper}

Much work in the study of quantum logic has been focused on the structure of the propositional lattice associated with a physical theory, which in \cite{Yousef:CoeventsAsBeables} we refer to as the \emph{logical structure}. Sorkin's coevent approach \cite{Sorkin:coeventsI,Sorkin:coeventsII} shifts the focus onto truth valuations; in particular the `non-classical' nature of a physical theory is manifested in the anhomomorphism of such valuations rather than (or in addition to) the non-Boolean structure of the propositional lattice. This leads us to extend the idea of a logical structure to that of a \emph{logical framework}; associating with the propositional lattice $\EA$ a truth value space $\T$ and a set of `allowed' truth valuation functions $\V$ from the lattice to this space, yielding a triple $(\EA,\V,\T)$ \cite{Yousef:CoeventsAsBeables}.

In this paper we will apply this focus on truth valuations to the consistent histories approach. In section $1.3$ we begin by reviewing the histories approach, in section $1.4$ we consider interpretation, in particular consistent histories, Isham's varying set construction for consistent histories \cite{Isham:1997} and Sorkin's coevents \cite{Sorkin:coeventsI,Sorkin:coeventsII}. In section $2.1$ we rephrase the consistent histories approach in terms of truth valuations and develop a logical framework to describe them. In section $2.2$ we consider how this might relate to Sorkin's coevent interpretation, particularly the multiplicative scheme, while in section $2.3$ we adapt Isham's varying set construction to our valuation based approach, building a new logical structure. We conclude in section $3$.

\subsection{The Histories Approach}

\subsubsection{Quantum Measure Theory}\label{subsubsec:QMT}

Sorkin's quantum measure theory \cite{Sorkin:1994dt,Sorkin:1995nj} formulates the histories approach as a generalisation of probability theory. A \emph{measure theory} or \emph{histories theory} is a triple $\H$ in which $\S$ is the \emph{sample space}, $\EA$ the \emph{event algebra} and $\mu$ the \emph{measure}. The elements of the sample space are the maximally detailed descriptions of outcomes for the system under examination, for example if we are considering a two tosses of a standard coin we have $\S=\{hh,ht,th,tt\}$ where $h=heads$ and $t=tails$, alternately for describing the dynamics of a particle we might take $\S$ to be the set of spacetime paths from a set starting point. The event algebra $\EA\subseteq 2^{\S}$ is our propositional lattice, with the lattice structure inherited from that of $2^{\S}$, which in turn is defined by set inclusion. The measure $\mu:\EA\rightarrow\R$ describes the dynamics, we require that $\mu(A)\geq 0$ and $\mu(\S)=1$. If $\mu$ obeys the Kolmogorov sum rule,
\beq
\mu(A\sqcup B)=\mu(A)+\mu(B) \ \forall \ A,B\in\EA,
\eeq
where $\sqcup$ denotes disjoint union, then it is a probability measure and $\H$ a probability theory. If the dynamics is already described by a decoherence functional\footnote{The decoherence functional $D:\EA\times\EA\rightarrow\C$ is the standard form of the dynamics in a histories theory, for more details see for example \cite{Hartle:1992as}} $D$ we set $\mu(A)=D(A,A)$ $\forall A\in\EA$. In general this leads $\mu$ to obey the \emph{quantum sum rule},
\bea
\mu(A\sqcup B\sqcup C) &=& \mu(A\sqcup B)+\mu(B\sqcup C)+\mu(C\sqcup A) \nn \\
&& -\mu(A)-\mu(B)-\mu(C), \ \ \forall A,B,C\in\EA \label{eq:level 2 sum rule}
\eea
We will call $\mu$ a \emph{quantum measure} if it obeys (\ref{eq:level 2 sum rule}), in which case we will denote $\H$ a \emph{quantum measure theory} or simply a \emph{quantum theory}. Notice that a probability measure is a quantum measure, so that probability theories are a subset of quantum theories.

Unless explicitly stated otherwise, in what follows we will work with quantum measure theories, and for simplicity will assume that the sample space is finite and that $\EA=2^{\S}$, so that the event algebra is Boolean\footnote{In what follows we will use the terms `Boolean algebra' and `Boolean lattice' interchangeably.}.

\subsubsection{Partitions and Coarse Grainings}\label{subsubsec:partitions}

We refer to a partition $\Ll$ of $\S$ as a \emph{coarse graining}; $\Ll$ generates the \emph{coarse grained Boolean sublattice} $\EAL$ of $\EA$ to which we can restrict the measure leading to the \emph{coarse grained histories theory} $\HL$. Further, we refer to $\Omega$, $\EA$ and $\H$ as \emph{fine grainings} of $\Ll$, $\EAL$ and $\HL$ respectively. We follow \cite{Isham:1997} in denoting the set of these coarse grained Boolean lattices\footnote{Note that \cite{Isham:1997} uses $\B$ to refer to the poset of \emph{all} Boolean sublattices, whereas we require such algebras to be generated by a partition.} by $\B$, which we can make into a poset using inclusion; $\EALa\leq \EALaa$ iff $\EALa\supseteq \EALaa$, so that $\EALa$ is a fine graining of $\EALaa$. Now in some cases the restriction of the measure to a particular subalgebra $\EAL$ might obey the Kolmogorov sum rule, so that the coarse grained theory $\HL$ is a probability theory. We call such subalgebras (and their associated partitions) \emph{dynamically classical} or \emph{decoherent}, and denote the poset of such algebras (ordered by inclusion) by $\B_D$. Note that the decoherence of $\EAL\in\B$ implies the decoherence of all coarse grainings of $\EAL$, so that $\B_D\subseteq \B$ is an upper set.

We may want to distinguish between decoherent partitions and the subset thereof which, at least in principle, correspond to observable alternatives. Now the question of what constitutes a physically observable event may be difficult to define appropriately, and is not pertinent to what follows; we simply wish to avoid the assumption that it coincides with inclusion in a decoherent set. Then without exploring how we might concretely define this concept for a general histories theory, we will label the poset of \emph{observable} Boolean sublattices by $\B_O\subseteq\B_D$, and will simply require that $\B_O$, like $\B_D$, should be an upper set. Continuing on this theme, we may wish to pick out those partitions corresponding to experiments that have actually taken place, or observations that have actually been made. Again, without going into what exactly we mean by an experiment or an observation in a closed system  we will denote the corresponding poset by $\B_E$ and require that it too should be an upper set.

\subsection{Interpretation}

\subsubsection{Consistent Histories in Quantum Measure Theory}

The interpretation of a probability theory is better understood than the interpretation of a quantum theory; the consistent histories approach \cite{Griffiths:1984rx,Griffiths:1993,Griffiths:1996,Griffiths:1998} therefore seeks to restrict the interpretation of a quantum measure theory to the subsystems which are (dynamically) classical, in other words to the elements of $\B_D$. Then in place of a single propositional lattice $\EA$ we will have a plethora of propositional lattices, $\EAL\in\B_D$. The propositions within any element of $\B_D$ can be logically compared using the full toolbox of Boolean algebra, however propositions which are not contained within a common element of $\B_D$ are deemed incomparable. In other words we keep the Boolean algebras $\EAL\in\B_D$ but `throw away' or `choose to ignore' the larger algebra $\EA$ (unless of course $\EA$ is itself decoherent). The decoherent subalgebras are often referred to as \emph{consistent sets}.

As an aside, note that in place of $\B_D$ we might employ some concrete realisation of $\B_O$ or $\B_E$. In particular, the use of $\B_E$ might perhaps be considered as the consistent histories approach to an instrumentalist interpretation.

\subsubsection{A Topos interpretation of Consistent Histories}\label{subsubsec:topos for consistent histories}

The separation of propositions into incomparable sets may itself require some interpretation. Isham addresses this issue by using topos theoretic structures to bring the various consistent sets together into a single logical formalism \cite{Isham:1997}; this work may be considered as a stand alone construction independent of the later and more far reaching application of topos theory by Isham and D\"{o}ring \cite{Isham:toposI,Isham:toposII,Isham:toposIII,Isham:toposIV}. We will briefly outline some key results of \cite{Isham:1997} which we will need further on, the reader is referred to \cite{ToposBook} for a more detailed description and an introduction to the relevant aspects of topos theory. Starting with a histories theory $\H$ we proceed in steps.
\begin{enumerate}
    \item {We construct the constant varying set $\triangle\EA$ in the topos $\mathbf{Sets^{\B_D}}$, which associates $\EA$ with each $\EAL\in\B_D$.}
    \item {Note that the subobject classifier $\sO$ in this topos is the association with each $\EAL\in\B_D$ of the set of upper sets in $\B_D$ whose elements are all greater than or equal to $\EAL$.}
    \item {We construct the subobject $S_D$ which associates with each $\EAL\in\B_D$ a set of propositions which Isham refers to as \emph{accessible} from $\EAL$ \cite{Isham:1997},
    \beq
    S_D(\EAL) = \{A \ | \ \exists \EAP\leq \EAL \ \text{st} \ A\in\EAP \}.
    \eeq
    Essentially this means that $A$ is an element of at least one fine graining of $\EAL$, or of $\EAL$ itself.}
    \item{We define the maps,
    \bea
    \chi_{\EAL}^{S_D}:\triangle\EA(\EAL)&\rightarrow& \sO(\EAL) \nn \\
    \chi_{\EAL}^{S_D}:A&\mapsto& \{\EAP\geq\EAL \ | \ A\in S_D(\EAL)\}.
    \eea
    Putting together $\chi_{\EAL}^{S_D}$ gives us the characteristic morphism $\chi^{S_D}:\triangle\EA\rightarrow\sO$.}
    \item{Given any event that is a member of a decoherent algebra, $A\in\EAL\in\B_D$, we can use the characteristic morphism to define a corresponding global element of the subobject classifier,
        \beq
        \G\langle A\rangle: \mathbf{1}\rightarrow\sO,
        \eeq
        where $\mathbf{1}$ is the terminal object in $\mathbf{Sets^{\B_D}}$, which corresponds to the association of a singleton set $\{\bullet\}$ with each $\EAL\in\B_D$ (any singleton set will suffice up to isomorphism). Then as a morphism $\G\langle A \rangle$ is defined by the family of maps,
        \beq
        \G\langle A\rangle_{\EAL}:\bullet\mapsto\chi_{\EAL}^{S_D}(A).
        \eeq
        }
    \item{We can use these global elements $\G\langle A\rangle$ to bring together the various propositional lattices $\EAL\in\B_D$ into a single propositional lattice. The subobject classifier is possessed of a natural Heyting algebra structure, so the set,
        \beq
         \G\langle \EA\rangle = \{ \G\langle A\rangle \ | \ \exists\EAL\in\B_D \ \text{st} \ A\in\EAL\},
        \eeq
        can be thought of as a subset of a Heyting algebra, and so will generate a Heyting algebra $ H\langle \EA\rangle$. The reader is referred to \cite{Isham:1997} for further details.}
\end{enumerate}

\subsubsection{Truth Valuations and a Logical Framework for Probability Theories}

Rather than simply focusing on the structure of the propositional lattice, Sorkin has emphasised the role of truth valuations thereon. In \cite{Sorkin:coeventsII} Sorkin alludes to the `threefold character' of logic; in \cite{Yousef:CoeventsAsBeables} we formalised this notion to a triple $(\EA,\V,\T)$ which we called a \emph{logical framework}. The \emph{domain} is the event algebra $\EA$, the space of \emph{truth values} $\T$ is generally (but not necessarily) a lattice, and the space of \emph{truth valuations} is the set of `allowed' valuation maps $\f:\EA\rightarrow\T$. Note that strictly speaking, given $\V$ both $\EA$ and $\T$ are redundant since the definition of a map specifies its domain and range; however we will continue to explicitly include them for emphasis.

When $\H$ is a probability theory we typically use $\T=\Z2$, where we interpret $0$ as `false' and $1$ as `true'; notice that $\Z2$ is a Boolean lattice under the natural ordering $0<1$. The usual interpretation of a probability theory is that `one history occurs'; labelling this \emph{real history} as $r$ we can construct a valuation map,
\bea
r^*:\EA&\rightarrow&\Z2 \nn \\
r^*(A) &=& \left\{\begin{array}{cc} 1 & r\in A \\ 0 & r\not\in A \end{array}\right.
\eea
It is easy to check that such maps are lattice homomorphisms $r^*\in Hom(\EA,\Z2)$, in fact they are all of the lattice homomorphisms other than the zero map \cite{Yousef:thesis}, which we will exclude from $Hom(\EA,\Z2)$. Then for a classical probability theory $\HP$ our logical framework will be $(\EA,Hom(\EA,\Z2),\Z2)$.

Now we may want to link our logical framework to the dynamics in some way; perhaps the simplest means of doing so is Sorkin's concept of \emph{preclusion} \cite{Sorkin:coeventsI}, which intuitively states that dynamically `disallowed' events are logically `false'. More precisely this means that for a histories theory $\H$ we wish to impose,
\beq\label{eq:preclusion}
\mu(A)=0 \Rightarrow \f(A)=0 \ \forall A\in\EA, \f\in\V.
\eeq
This leads us to define,
\beq
\Cl=\{\f\in Hom(\EA,\Z2) \ | \ \mu(A)=0 \Rightarrow \f(A)=0\},
\eeq
so that when $\mu$ is a probability measure we can use the logical framework $(\EA,\Cl,\Z2)$.

\subsubsection{Coevents and the Multiplicative Scheme}\label{subsubsec:coevents}

It has been shown \cite{Yousef:KSP,Yousef:thesis} that there are gedanken experimentally realisable quantum theories for which $\Cl=\emptyset$, leading us to conclude that $(\EA,\Cl,\Z2)$ is not in general a suitable logical framework for a quantum histories theory. Sorkin's \emph{coevents} \cite{Sorkin:coeventsI,Sorkin:coeventsII} are a proposed generalisation in which the valuation maps may be anhomomorphic. More precisely, given a histories theory $\H$ a coevent is a map,
\beq
\f:\EA\rightarrow \Z2,
\eeq
which is not everywhere zero; we denote the set of all coevents by $\EA^{\diamond}$. A \emph{coevent scheme} is an assignment of a subset of $\EA^{\diamond}$ to every histories theory $\H$; we think of a coevent scheme as representing the `physically allowed' set of truth valuations for each histories theory.

A coevent is \emph{multiplicative} \cite{sorkin:coeventsII} if it obeys,
\beq\label{eq:multiplicative rule}
\f(A\wedge B) = \f(A)\wedge\f(B).
\eeq
We denote the set of multiplicative coevents associated with a histories theory $\H$ by $\MCE$. It is easy to see that the support $\f^{-1}(1)$ of a multiplicative coevent is a filter. Since $\EA$ is finite, we can define the map,
\bea
*:\EA^*&\rightarrow&\EA \nn \\
*:\f&\mapsto&\f^*,
\eea
where $\f^*$ is the principal element of $\f^{-1}(1)$. We can extend this map to $\EA\sqcup\EA^*$ by setting $*:A\mapsto A^*$ for all $A\in\EA$ where,
\beq
A^*(B)=\left\{\begin{array}{cc} 1 & A\subseteq B \\ 0 & A\not\subseteq B \end{array}\right.
\eeq
It is easy to see that $(\f^*)^*=\f$, so that $*$ is an involution; because of this we can think of $*$ as a duality. Comparison of multiplicative coevents $A^*$ with homomorphisms $r^*$ has lead to multiplicative coevents being considered as `\emph{ontological coarse grainings}'.

We say that a coevent is \emph{preclusive} if it obeys (\ref{eq:preclusion}), and that a coevent $\f$ \emph{dominates} a coevent $\psi$ if $\f(A)=1\Rightarrow\psi(A)=1 \ \forall A\in\EA$. A coevent $\f$ is \emph{minimal} within a set $S\subset\EA^{\diamond}$ of coevents if there does not exist $\psi\in S$ such that $\psi$ dominates $\f$. The \emph{multiplicative scheme} associates with a histories theory $\H$ the set of coevents that are minimal among preclusive multiplicative coevents \cite{Sorkin:coeventsII,Yousef:thesis}.

The purpose of minimality is to ensure that $\M=\Cl$ whenever the measure is classical (whenever $\mu$ is a probability measure), so that a dynamically classical system will obey `classical logic'. However in practise we expect probability theories to `emerge' as the restriction of a quantum theory to a decoherent sublattice, which suggests that we should look for some kind of `coevent classicality' on such sublattices $\EAL\in\B_D$. At the very least, to avoid conflict with observation we require some form of `classicality' in $\B_E$. To make this more precise, we say that a coevent $\f$ is \emph{classical} on a partition $\Ll$ (or the boolean lattice $\EAL$ generated by $\Ll$) if $\f$ maps exactly one of the elements of $\Ll$ to one, which is equivalent to $\f$ restricting to a (not everywhere zero) homomorphism on $\EAL$.

Now it can be shown that coevents $\f\in\M$ in the multiplicative scheme are not in general classical on all decoherent partitions, leading to potential interpretational difficulties. Sorkin \cite{Sorkin:coeventsII} has suggested an alternate dynamical condition, that of \emph{preclusive separability}. A partition $\{A_i\}$ (and the associated sublattice $\EA_{\{A_i\}}$) is preclusively separable for any measure zero set $Z$ the intersections $A_i\cap Z$ are also of measure zero, for all $i$. We will write $\B_P$ to denote the poset of preclusively separable sublattices, it can me checked that $\B_P$ is an upper set. It can be shown \cite{Yousef:thesis} that any $\f\in\M$ is classical on any preclusively separable partition; further, for any preclusive homomorphism $\psi\in\ClL$ on a preclusively separable partition $\Ll$ there exists a $\f\in\M$ which restricts to $\psi$ on $\EAL$. Sorkin has argued \cite{Sorkin:coeventsII,Sorkin:private} that any partition relating to a potential observation is preclusively separable, which in our terminology means $\B_P\subseteq\B_O$.

\section{Consistent Histories as Truth Valuations}

\subsection{A Logical Framework for Consistent Sets}\label{subsec:consistent histories logical framework}

Consistent histories focuses on the individual consistent sets, which in our terminology are the dynamically classical sublattices. For any such $\EAL\in\B_D$ the dynamics restricts to a probability theory $\HL$; the consistent histories approach directs us to interpret this restricted theory on a `stand alone' basis. Following the above, we then construct the logical framework $(\EAL,Hom(\EAL,\Z2),\Z2)$ to describe the system, alternatively we might use $(\EAL,\ClL,\Z2)$ if we wish to rule out dynamically precluded valuations.

This leaves us with a logical framework for every dynamically classical partition. In section \ref{subsection:topos for valuations} we will see how the partial order of $\B_D$ can be used to bring these frameworks into a single structure along the lines of Isham's topos approach to consistent histories \cite{Isham:1997}. However for now we will simply bring all the valuations together by generalising our definition of a logical framework to be a triple $\LlF$, where $\D$ is a set of sublattices of $\EA$ and each of the valuations in $\V$ now maps from \emph{one} of the elements of $\D$ to $\T$. Then if $\H$ is a probability theory the relevant logical framework would be $(\{\EA\},\Cl,\Z2)$, and if $\H$ is a quantum theory the multiplicative scheme would yield $(\{\EA\},\M,\Z2)$. Returning to consistent histories approach, we therefore suggest the logical framework $(\B_D,\V_D,\Z2)$ where,
\beq
\V_D=\bigcup_{\EAL\in\B_D}Hom(\EAL,\Z2).
\eeq
Alternatively if wish to rule out the dynamically precluded valuations at this stage we would use $(\B_D,\V_C,\Z2)$ where,
\beq
\V_C=\bigcup_{\EAL\in\B_D}\ClL.
\eeq

\subsection{Comparison with Multiplicative Coevents}\label{subsec:comparison with mult coevents}

We label by $\fla$ the valuation in $Hom(\EAL,\Z2)$ mapping $A\in\Ll$ to $1$; note that $\fla$ is unique and $\fla(B)=0$ for all $B\in\Ll$ such that $B\neq A$. Note also that the `partition index' $\Ll$ is necessary; $A^*_{\Ll_1}\neq A^*_{\Ll_2}$ since the two maps have different domains.

Now notice that,
\beq
supp(\fla)=supp(A^*),
\eeq
where $supp(\f)=\f^{-1}(1)$ is the support of a $\Z2$ valued map $\f$. In fact $\fla$ is simply the restriction of $A^*$ to the sublattice $\EAL$. This suggests a close link between consistent histories and multiplicative coevents; we will examine two approaches to bringing them together.
\begin{description}
    \item[1. Adjust the multiplicative scheme to accommodate consistent histories]
    {Given a histories theory $\H$ we could use the consistent histories logical framework to define a new coevent scheme. The most obvious way to do this would be by extending $\fla$ to the multiplicative coevent $A^*$. This would yield the coevent scheme,
    \beq
    Cons_D \H = \{A^* \ | \ \exists \fla\in\V_D \ \text{st} \ supp(A^*)=supp(A^*_{\Ll}) \}.
    \eeq
    Notice that $Cons_D\H\subseteq\EA^*$ so that `consistent histories coevents' are multiplicative. However they will not in general be preclusive; to remedy this we can define,
    \beq
    Cons_C \H = \{A^* \ | \ \exists \fla\in\V_C \ \text{st} \ supp(A^*)=supp(A^*_{\Ll}) \}.
    \eeq
    Note that if $\exists Z\in\EA$ st $\mu(Z)=0$ and $A^*(Z)=1$ then,
    \bea
    && Z\in supp(A^*) \nn \\
    &\Rightarrow& Z\in supp(\fla) \nn \\
    &\Rightarrow& Z\in\EAL.
    \eea
    But then we have $Z\in\EAL$ with $\mu |_{\Ll}(Z)=0$ and $\fla(Z)=1$, so $\fla\not\in\ClL$. Therefore any $A^*\in  Cons_C \H$ is preclusive.

    We might also choose to impose a minimality condition, for example we could start with the set $M_{PC}$ of all multiplicative coevents that are both preclusive and classical on every decoherent partition, and then choose the minimal elements\footnote{The coevents that are not dominated by any other coevent in $M_{PC}$, see section \ref{subsubsec:coevents}.} of $M_{PC}$. We refer to these coevents as $Cons_M\H$.

    Now $Cons_C\H$ and $Cons_M\H$ could be thought of as alternatives to (or adjustments of) the multiplicative scheme $\M$. However in general coevents in $Cons_C\H$ or $Cons_M\H$ will not be classical on \emph{all} classical partitions, and unlike the case of the multiplicative scheme we can not even argue that they will be classical on observable lattices $\EAL\in\B_O$. Further, we have no guarantee that $Cons_M\H$ will not be empty.
    }
    \item[2. Adjust consistent histories to accommodate the multiplicative scheme]
    {Perhaps the most elegant means of achieving this is to use $\B_{PD}=\B_P\cap\B_D$ in place of $\B_D$, requiring a sublattice $\EAL$ to be \emph{both} decoherent and preclusively separable if it is to be considered as a `consistent set'. We would then construct the logical framework $(\B_{PD},\V_{PD},\Z2)$ where,
    \beq
    \V_{PD} = \bigcup_{\EAL\in\B_{PD}} Hom(\EAL,\Z2),
    \eeq
    or the preclusive counterpart,
    \beq
    \V_{PD} = \bigcup_{\EAL\in\B_{PD}} \ClL.
    \eeq

    In this framework, our results on preclusive separability (section \ref{subsubsec:coevents} and \cite{Yousef:thesis}) mean that any $\f\in\M$ will be classical on any `consistent set' $\EAL\in\B_{PD}$. Furthermore any $\fla\in\V_{PD}$ will be the restriction of some $\f\in\M$ to $\EAL$. This effectively combines the multiplicative scheme with our consistent histories logical framework. However this will only be of interest if we accept Sorkin's argument \cite{Sorkin:coeventsII} that $\B_P\subseteq\B_O$, and only if we are happy with `non-classical logic' on decoherent but not preclusively separable lattices.
    }
\end{description}

\subsection{Varying Sets for Valuations}\label{subsection:topos for valuations}

We can adapt Isham's treatment of consistent histories (section \ref{subsubsec:topos for consistent histories} and \cite{Isham:1997}) to our valuation version thereof. This leads us to use the poset structure of $\B_D$ to bring together the logical frameworks $(\EAL,Hom(\EAL,\Z2),\Z2)$ into a single construction that may be viewed as an alternative to $(\B_D,\V_D,\Z2)$. Alternatively we might view this as a process of adding structure to $(\B_D,\V_D,\Z2)$ based on the order relations within $\B_D$.

If we were to follow Isham's approach exactly for valuations, we would associate with each element of $\B_D$ the whole event algebra $\EA$ to yield a constant varying set which we label $\triangle^*\EA$. We would then form the subobject $S^*_D$ where,
\beq
S^*_D(\EAL)=\{A^* \ | \ A\in S_D(\EAL) \},
\eeq
from whence we can construct a valuation morphism and proceed as in section \ref{subsubsec:topos for consistent histories}; due to the duality between $\EA$ and $\EA^*$ the resulting structures would be isomorphic to those of section \ref{subsubsec:topos for consistent histories}. However, the constant varying set $\triangle^*\EA$ seems intuitively to be more like a description of multiplicative coevents than of our consistent histories valuations. We therefore refine this construction to better fit consistent histories.

Instead of using a constant varying set, we will simply associate with each $\EAL\in\B_D$ the valuations arising from the corresponding partition $\Ll$. First we note that for $\EALa,\EALaa\in\B_D$ we have,
\beq
\EALa \leq \EALaa \Leftrightarrow \EALa \supseteq \EALaa.
\eeq
It then can be checked that every map $\f_2\in Hom(\EALaa,\Z2)$ is the restriction to $\EALaa$ of one or more maps $\f_1\in Hom(\EALa,\Z2)$. Now although there may be many $\f_1\in Hom(\EALa,\Z2)$ corresponding to a given $\f_2\in Hom(\EALaa,\Z2)$, the restriction of $\f_1\in Hom(\EALa,\Z2)$ to $\EALaa$ is unique. This allows us to define the map,
\bea
V_{D12}:Hom(\EALa,\Z2)&\rightarrow& Hom(\EALaa,\Z2) \nn \\
V_{D12}(\f_1) &=& \f_1 |_{\EALaa}.
\eea
Then composing such maps, and writing $\V_{Di}=Hom(\EALi,\Z2)$, for any pair $\EALi\leq\EALj$ we can define,
\bea
V_{Dij}:\V_{Di} &\rightarrow& \V_{Dj} \nn \\
V_{Dij}(\f_i) &=& \f_i |_{\EALj}.
\eea
It can be checked that the assignment of $\V_{Di}$ to each $\EALi\in\B_D$, together with the maps $V_{Dij}$, forms a \emph{varying set over $\B_D$} which we will denote $\vV$. Note that $\vV$ is not a constant varying set.

Intuitively, the `dual' of $\vV$ would be the varying set associating the elements of the partition $\Ll$ to each $\EAL\in\B_D$, together with the map sending $A\in\Ll$ to the unique $B\in\Pi$ satisfying $B\supseteq A$, where $\Pi$ is a coarse graining of $\Ll$ (so that $\EAP\geq\EAL$). However this structure seems less natural than $\vV$.

We can now adapt the subobject $S^*_D$ to our new construct. Because $\triangle^*\EA$ associates $\EA^*$ to each $\EAL\in\B_D$ whereas $\vV$ associates $Hom(\EAL\Z2)\subseteq\EA^*$ to $\EAL$, it is therefore natural to consider a `restriction' of $S^*_D$. With this in mind we define the object $\sV$ in $\mathbf{Sets^{\B_D}}$,
\beq
\sV(\EAL)=\{\fpa |_{\EAL} \ | \ \exists\EAP\leq\EAL \ \text{st} \ \fpa\in Hom(\EAP,\Z2) \},
\eeq
together with the maps,
\bea
\sV_{ij}:\sV(\EALi)&\rightarrow&\sV(\EALj) \nn \\
\sV_{ij}:\f_i&\mapsto&\f_i |_{\EALj},
\eea
for $\EALi \leq \EALj$. We check that $\sV$ is well defined; firstly it is easy to see that $\EAP\leq\EAL\Rightarrow \fpa |_{\EAL}\in Hom(\EAL,\Z2)=\vV(\EAL)$. Secondly, we have,
\beq
\sV_{ij}:\sV(\EALi) \mapsto \{ (\fpa |_{\EALi})|_{\EALj} \ | \ \exists\EAP\leq\EALi \ \text{st} \ \fpa\in Hom(\EAP,\Z2) \}.
\eeq
Since $\EAP\supseteq\EALi\supseteq\EALj$ we have $(\fpa |_{\EALi})|_{\EALj}=\fpa|_{\EALj}$, so that,
\bea
\sV_{ij}:\sV(\EALi) &\mapsto& \{ \fpa |_{\EALj} \ | \ \exists\EAP\leq\EALi \ \text{st} \ \fpa\in Hom(\EAP,\Z2) \} \nn \\
&\subseteq& \{ \fpa |_{\EALj} \ | \ \exists\EAP\leq\EALj \ \text{st} \ \fpa\in Hom(\EAP,\Z2) \} \nn \\
&=& \sV(\EALj).
\eea
Thus $\sV$ is a well defined object in $\mathbf{Sets^{\B_D}}$.

It is also easy to check that $\sV$ is a subobject of $\vV$. Firstly note that,
\beq
\sV(\EAL)\subseteq\vV(\EAL).
\eeq
Secondly we see that $\sV_{ij}$ is simply a restriction of $\vV_{ij}$ to $\sV$, by which we mean that for $\f_i\in\sV(\EALi)$,
\bea
\vV_{ij}(\f_i) &=& \f_i |_{\EALj} \nn \\
&=& \sV_{ij}(\f_j).
\eea
In other words, the following diagram commutes for $\EALi\leq\EALj$,
\beq
\begin{CD}
 {\sV(\EALi)} @> {\sV_{ij}} >> {\sV(\EALj)} \\
 @ VVV @  VVV \\
 {\vV(\EALi)} @>> {\vV_{ij}} > {\vV(\EALj)} \\
\end{CD}
\eeq
where the vertical arrows are subset inclusions. Thus $\sV$ is indeed a subobject of $\vV$.

We can then follow \cite{Isham:1997} to define a characteristic map $\chi^{\sV}:\vV\rightarrow\sO$ as described in section \ref{subsubsec:topos for consistent histories},
\bea
\chi^{\sV}_{\EAL}:\vV(\EAL) & \rightarrow & \sO(\EAL) \nn \\
\f & \mapsto & \{\EAP\geq\EAL \ \ | \ \ \f |_{\EAP} \in\sV(\EAP)\},
\eea
which we can use to define (not necessarily distinct) global elements,
\beq
\G\langle \f \rangle :\mathbf{1}\rightarrow\sO,
\eeq
for each consistent histories valuation $\f\in\V_D$. Then we can use the Heyting algebra structure of $\sO$ to define the `lattice' logical operations of meet and join on the set,
\beq
\G\langle V_D \rangle = \{\G\langle \f \rangle | \f\in\V_D\}.
\eeq
The closure of meet and join on $\G\langle V_D \rangle$ will be a Heyting algebra \cite{Isham:1997}, which we call $H_D\langle V_D \rangle$. The elements of $H_D\langle V_D \rangle$ might best be thought of as propositions concerning the dynamics $\mu$ of the underlying histories theory; in particular the `structure' of decoherence implied by $\mu$. We have defined a map,
\bea
h_D:V_D&\rightarrow&H\langle V_D \rangle \nn \\
\f &\mapsto&\G\langle \f \rangle,
\eea
which effectively `embeds' the consistent histories valuations in a single logical structure.

It would of course be possible to replace the poset $\B_D$ with $\B_O$ or $\B_E$ to yield varying sets within the topoi $\mathbf{Sets^{\B_O}}$ or $\mathbf{Sets^{\B_E}}$, leading us to embed the corresponding sets of valuations in Heyting algebras $H_O\langle V_O \rangle$ and $H_E\langle V_E \rangle$ respectively. This might be thought of as an instrumentalist version of the above analysis. Alternately, we might use the poset $\B_{PD}$, which would lead to an accommodation between the consistent histories approach and the multiplicative scheme, as described in section \ref{subsec:comparison with mult coevents}.

Finally, we might seek to extend the logical structure $H_D\langle V_D \rangle$ to a logical framework. Noting that $\Z2$ is a Heyting algebra (as in every Boolean algebra), the most obvious starting point would be $(H_D\langle V_D \rangle,Hom(H_D\langle V_D \rangle,\Z2),\Z2)$; however we leave this investigation to future research.

\section{Summary and Conclusion}\label{sec:conclusion}

Our aim was to rephrase the consistent histories approach in terms of truth valuations, following Sorkin's development of the coevent interpretation. In section \ref{subsec:consistent histories logical framework} we identified each consistent set with a logical framework $(\EAL,Hom(\EAL,\Z2),\Z2)$, where $\Ll$ is the corresponding partition. We then generalised our definition of a logical framework to bring the consistent sets together as $(\B_D,\V_D,\Z2)$, or as $(\B_D,\V_C,\Z2)$ if we wish to take preclusion into account. In section \ref{subsec:comparison with mult coevents} we noted the close relation with multiplicative coevents given by $supp(\fla)=supp(A^*)$, and examined two means of combing the approaches. This led us to introduce the coevent schemes $Cons_D \H$, $Cons_C \H$ and $Cons_M \H$, as well as the logical framework $(\B_{PD},\V_{PD},\Z2)$. In section \ref{subsection:topos for valuations} we adapted Isham's early consistent histories topos construction to our valuations based approach, moving from the constant varying set $\triangle \EA$ to the varying set $\vV$ while remaining within the topos $\mathbf{Sets^{\B_D}}$. We hope that these results show that the consistent histories approach can be naturally described in terms of valuations, opening links with Sorkin's coevent interpretation and facilitating a more simple description in terms of Isham's varying set construction.

\newpage
\bibliography{Bib}
\bibliographystyle{plain}

\end{document}